\documentclass[conference]{IEEEtran}
\IEEEoverridecommandlockouts
\usepackage{cite}
\usepackage{amsmath,amssymb,amsfonts}
\usepackage{algorithmic}
\usepackage{graphicx}
\usepackage{textcomp}
\usepackage{xcolor}
\def\BibTeX{{\rm B\kern-.05em{\sc i\kern-.025em b}\kern-.08em
    T\kern-.1667em\lower.7ex\hbox{E}\kern-.125emX}}
    
\usepackage{algorithm} % for pseudo-code notation
\usepackage{algorithmic} % for pseudo-code notation
\usepackage{url} % for urls with underscores

\newcommand{\Unetwork}{\mathcal{G}}
\newcommand{\Unodes}{\mathcal{V}}
\newcommand{\Uedges}{\mathcal{E}}
\newcommand{\Uinneighbors}[1]{n_\text{down}(#1)}
\newcommand{\Uoutneighbors}[1]{n_\text{up}(#1)}
\newcommand{\dir}{d}
\newcommand{\R}{\mathbb{R}}
\newcommand{\embdim}{{N}}

\newcommand{\shortemb}{e}
\newcommand{\error}[2]{E(#1,#2)}
\newcommand{\shortpred}{\hat{e}}
\newcommand{\incoherence}{I}
\newcommand{\pivot}{P}
\newcommand{\coneighbors}{C}

\newcommand{\MPset}{\mathcal{P}}
\newcommand{\MPfol}{\mathcal{F}}
\newcommand{\MPfilfol}{\tilde{\mathcal{F}}}
\newcommand{\Aset}{\mathcal{A}}
\newcommand{\Bset}{\mathcal{B}}
\newcommand{\Cset}{\mathcal{C}}

\newcommand{\result}{\mathcal{R}}

\newcommand{\setPsize}{831}
\newcommand{\setFsize}{4.487.430} 
\newcommand{\setFtsize}{368.831}
\newcommand{\setAsize}{4.483}
\newcommand{\setBsize}{1.304.812}
\newcommand{\setCsize}{5.528.716}

\newcommand{\setFinterB}{138.424}
\newcommand{\setFinterC}{231.035}

\begin{document}

\title{Your most telling friends: \\Propagating latent ideological features on Twitter using neighborhood coherence
\thanks{This work has been funded by the French National Agency for Research under grant ANR-19-CE38-0006: Geometry of Public Issues (GOPI).}
}

\author{\IEEEauthorblockN{Pedro Ramaciotti Morales}
\IEEEauthorblockA{\textit{pedro.ramaciottimorales@sciencespo.fr}\\
\textit{m\'edialab, Sciences Po} \\
Paris, France \\
}
\and
\IEEEauthorblockN{Jean-Philippe Cointet}
\IEEEauthorblockA{\textit{jeanphilippe.cointet@sciencespo.fr}\\
\textit{m\'edialab, Sciences Po} \\
Paris, France \\
}
\and
\IEEEauthorblockN{Julio Laborde}
\IEEEauthorblockA{\textit{julio@recital.ai}\\
\textit{reciTAL} \\
Paris, France \\
}
}

\maketitle

\begin{abstract}
Multidimensional scaling in networks allows for the discovery of latent information about their structure by embedding nodes in some feature space.
Ideological scaling for users in social networks such as Twitter is an example, but similar settings can include diverse applications in other networks and even media platforms or e-commerce.
A growing literature of ideology scaling methods in social networks restricts the scaling procedure to nodes that provide interpretability of the feature space:
on Twitter, it is common to consider the sub-network of parliamentarians and their followers.
This allows to interpret inferred latent features as indices for ideology-related concepts inspecting the position of members of parliament.
While effective in inferring meaningful features, this is generally restrained to these sub-networks, limiting interesting applications such as country-wide measurement of polarization and its evolution.
We propose two methods to propagate ideological features 
beyond these sub-networks: one based on homophily (linked users have similar ideology), and the other on structural similarity (nodes with similar neighborhoods have similar ideologies).
In our methods, we leverage the concept of neighborhood ideological coherence as a parameter for propagation.
Using Twitter data, we produce an ideological scaling for 370K users, and analyze the two families of propagation methods on a population of 6.5M users.
We find that, when coherence is considered,
the ideology of a user is better estimated from those with similar neighborhoods, than from their immediate neighbors.

\end{abstract}


%%%%%%%%%%%%%%%%%%%%%%%%%%%%%%%%%%%%%%%%%%%%%%%%%%%%%%%%%%%%%%%%%%%%%%%%%%%%
\section{Introduction}
\label{sec:introduction}

Methods for embedding networks 
%in some feature space 
have become ubiquitous tools for analysis \cite{goyal2018graph,cai2018comprehensive}.
These methods exploit the geometrical representation of networks in a feature space,
which can be used in tasks such as compression \cite{wang2016structural}, clustering \cite{white2005spectral}, visualization \cite{pearson1901liii}, link prediction \cite{wang2016structural}, node classification \cite{tang2015line}, or scaling of a latent property that is determinant in the structure of networks.
Examples of this latter application can be found in the use of multidimensional scaling of networks for the retrieval of latent features in music for artist recommendation \cite{platt2004fast}, or in the use of Expectation Maximization algorithms for the estimation of ideological positions of voters, legislators, or online users \cite{imai2016fast}.
A growing domain of research has sought to exploit these %multidimensional 
scaling methods to infer ideological feature spaces in which to embed users of social networks.
Taking inspiration in ideological scaling of voting data, pioneered by Poole et al. in the 1980s \cite{poole1985spatial}, and developed in the 1990s into the widely-used NOMINATE method \cite{poole1991patterns}, some adaptations to social network data have been proposed. The first such adaptation was proposed by Bond et al. \cite{bond2015quantifying}, then popularized by Barbera in the mid- to late 2010s \cite{barbera2015birds,barbera2015tweeting}.
These new scaling methods have been successful in identifying latent features in social networks related to the ideology of users.
They are, however, often limited to small subsets of the whole network of users.
The reason for this may be found in computational limitations in computing scaling for large networks, but mostly in the need for interpretability for the found features.
On Twitter, traditionally,  an ideological scaling is limited to a sub-graph of parliamentarians and their followers, using the embedded features of the first ones to provide an interpretation for the found features \cite{barbera2015understanding}.
This raises the question:
How to compute features for portions of the network outside this \textit{seed} sub-graph?
The setting of this problem is similar to that of problems such as latent feature or link prediction, that accounts for a wealth of works to be discussed in Section~\ref{sec:related_work}.
But crucially, it deviates from these settings in that a minority of nodes have \textit{known} features (estimated via scaling), while those of a comparatively large set of nodes must be estimated from this initial seed set.
This marks a difference with statistical learning methods, where a flexible model is learned on a majority of nodes, and then applied to a minority of nodes with missing values.
The setting of ideological scaling invites the use of less flexible models, including strong assumptions about the link between ideology and the structure of the network.

This article takes inspiration in methods for information propagation in social networks and message coherence in telecommunications networks to propose a method for the propagation of scaled ideological features, from a seed set of nodes, to larger parts of a network.
While this problem arises in the case of ideological scaling in social networks in particular, it can also suit a more general class of problems in which n-dimensional features are only known for a subset of nodes in a network.
One may consider that latent ideological features are defined only for users that follow members of parliament (MPs).
However, our method supposes that a larger set of nodes could be positioned in the same latent feature space.
This underlying hypothesis is leveraged by our method exploiting two strong assumptions: 1) the ideology of a node is similar to that of its neighbors if this neighborhood is ideologically coherent, and 2) the ideology of a node is similar to other structurally similar nodes, that follow the same users for example, if these followed users are ideologically coherent.

After providing a review of the relevant related work in Section~\ref{sec:related_work}, we lay out the definitions and the notation needed for the treatment of the problem of coherent feature propagation in networks in Section~\ref{sec:preliminaries}.
Using this framework, we define the proposed propagation methods in Section~\ref{sec:value_propagation_methods}.
To analyze these methods, we develop an application case based on the ideology scaling on Twitter.
Section~\ref{sec:illustrative_case_data} presents the case of a seed set of Twitter accounts, followers of MPs in France, with known ideological scaling computed using procedures well-established in the literature.
Finally, in Section~\ref{sec:numerical_results}, we devise and execute experimental protocols to assess 1) the prediction capabilities    using both of our methods following the two hypotheses in estimating ideological features, and 2) the limits of the estimation in terms of the trade-off between precision of the estimation, and its reach as parts of the whole network.

%%%%%%%%%%%%%%%%%%%%%%%%%%%%%%%%%%%%%%%%%%%%%%%%%%%%%%%%%%%%%%%%%%%%%%%%%%%%
\section{Related Work}
\label{sec:related_work}

The methods proposed in this article are based in the propagation of features, from nodes for which they are known, to nodes for which they are not. This is mainly related to three different domains of research here reviewed. We intentionally omit the --vast-- scientific literature regarding the identification of important and central nodes in networks, outside the scope of these selected domains. 

%%%%%%%%%%%%%%%%%%%%%%%%%%%%%%%%%%%%%%%%%%%%%
\subsubsection{Latent ideological features in social networks}

The first category of related works include those that compute latent features in social networks.
While several such methods exist, this article is concerned by those that leverage the interpretability of a subset of nodes to extract meaningful feature spaces. 
This stands in contrast to works that produce feature embeddings on which to perform statistical learning such as deep learning, without necessarily providing an interpretation for the embedded features (cf. \cite{yan2006graph}).
A numerous family of methods, connected with link prediction, learn latent features in social networks, but specifically in relation with the probability of existence of a link, to be used in the propagation of local features into other parts of the time-evolving network \cite{heaukulani2013dynamic,wang2016structural}, or the retrieval of missing features for some nodes \cite{tang2015line}.
The most relevant works for this article, are those that use multidimensional scaling to extract ideological  features \cite{barbera2015birds,barbera2015tweeting}.
Other works seek to predict ideology for large networks, linking prediction to homophily, but use ideological labels and not a --continuous-- ideological feature space \cite{colleoni2014echo,himelboim2016valence,xiao2020timme}.

%%%%%%%%%%%%%%%%%%%%%%%%%%%%%%%%%%%%%%%%%%%%%
\subsubsection{Propagation in social networks}

A second category of works pertinent for this study concerns the propagation in social networks, accounting for numerous results in a long tradition, but organized here in propagation of \textit{information} and of \textit{properties}.
Propagation of information counts important cases such as the spread of rumors and misinformation \cite{bao2013new}, with connection to epidemiological models \cite{sahneh2014competitive,chakrabarti2008epidemic}.
This setting is fundamentally different from the one of this article in that it considers the network as the support for the flow of messages that have specific positions in time and space (as opposed to permanent latent features or properties).
However, there are possible connections between network structure and homophily on the one hand, and structure in networks and flow of information on the other \cite{aral2009distinguishing,mcpherson2001birds}.
This connection is also related, for example, to types of users in rumor propagation cascades in social networks \cite{friggeri2014rumor}.

Finally, propagation in social networks has connections with label propagation on graphs, used for example in heuristic procedures in community detection \cite{zhu2002learning,cheng2010modeling}. However, again, this differs from the setting of this article in that a label can be equated to discrete classification, whereas features can provide (n-dimensional) continuous indices for nodes.

%%%%%%%%%%%%%%%%%%%%%%%%%%%%%%%%%%%%%%%%%%%%%
\subsubsection{Node coherence}

Node coherence is a concept that can be encountered in domains such as neuroscience \cite{thagard1998coherence},  artificial intelligence \cite{mackworth1977consistency}, or communications networks \cite{larsson2004large}, and often addresses the degree of dissimilarity of signals coming from different sources.
This concept is of particular importance in multi-path networks.
Depending on the type of network, coherence can be taken into account for node properties rather than relayed messages.
The most relevant domain of works along this line, is found in node coherence in opinion dynamics, where an opinion is often modeled as a continuous feature \cite{edmonds2012modeling}.
In this domain, the concept of opinion coherence is sometimes considered in dynamic models \cite{rodriguez2016collective}.
In opinion dynamics, however, the focus is put in measuring different indices for node coherence to then be used in simulations to study evolution of features in systems.

%%%%%%%%%%%%%%%%%%%%%%%%%%%%%%%%%%%%%%%%%%%%%%%%%%%%%%%%%%%%%%%%%%%%%%%%%%%%
\section{Preliminaries}
\label{sec:preliminaries}

This section proposes the notation and definitions required to treat the concept of node and neighborhood coherence in networks, before proposing feature propagation methods in Section~\ref{sec:value_propagation_methods}.

%%%%%%%%%%%%%%%%%%%
\subsection{The universe network}
\label{sec:universe_network}

Let us consider a large \textit{universe network} as a directed graph $\Unetwork = (\Unodes,\Uedges)$, for some non-empty set $\Unodes$ with directed edges $\Uedges\subseteq\Unodes\times\Unodes$.
In social networks, nodes typically represent user accounts, and directed edges represent a relation of \textit{following} between them.
Information flows \textit{downstream}, contrary to the direction of the edges: if a user shares information, it will be received by her followers and not by her  \textit{followees}\footnote{Followees are also called \textit{friends} on Twitter}.

\begin{figure}[h!]
    \centering
    \includegraphics[width=0.5\columnwidth]{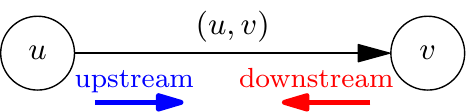}
    \caption{Convention used for \textit{upstream} and \textit{downstream} directions on directed social networks, where edges indicate following/friendship.}
    \label{fig:information_flow}
\end{figure}

For a node $v\in\Unodes$, we consider the set of its \textit{downstream} or in-neighbors $\Uinneighbors{v}=\left\{ u \in \Unodes : (u,v)\in\Uedges \right\}$ and its \textit{upstream} or out-neighbors $\Uoutneighbors{v}=\left\{ u \in \Unodes : (v,u)\in\Uedges \right\}$. Abusing notation,  we define the \textit{downstream} neighborhood of a set of nodes $V\subset\Unodes$ as $\Uinneighbors{V} = \left\{ u \in \Unodes : \exists v\in V \left( (u,v)\in \Uedges \right) \right\}$, 
and its \textit{upstream} neighborhood as $\Uoutneighbors{V} = \left\{ u \in \Unodes : \exists v \in V \left( (v,u)\in \Uedges \right) \right\}$. %\JL{same thing here}
It is worth noticing that, in general, we cannot assure $V\cap n_\dir (V)=\varnothing$, for $\dir \in\{\text{up},\text{down}\}$.
Given a direction $\dir$, we will denote its opposite direction with a bar, as $\overline{\dir}$.

%%%%%%%%%%%%%%%%%%%%%%%%%%%%%%%%%%%%%%%%%
\subsection{Features of nodes and estimation error}
\label{subsec:nodes_features}

Let us suppose that we can attribute $\embdim$-dimensional features for some limited set of nodes $V\subset\Unodes$.
When available for a node $v\in\Unodes$, we denote their known $\embdim$-dimensional features by $\shortemb(v)\in\R^\embdim$, and its $i$-th component by $\shortemb_i (v)$.
When initially unavailable for a node $v\in\Unodes$, $\embdim$-dimensional features may be attributed through feature propagation from nodes for which they are known.
Methods for doing so are at the core of this article, and will be proposed in Section~\ref{sec:value_propagation_methods}.
We denote estimated features for a node $v\in\Unodes$ by $\shortpred(v)\in\R^\embdim$, and by $\shortpred_i(v)$ its $i$-th component.

Whenever we can estimate features $\shortpred(v)$ through propagation methods for a node $v\in\Unodes$ for which we also know its \textit{true} features $\shortemb(v)$, we can compute the estimation error $\error{\shortpred(v)}{\shortemb(v)}$.
We measure this estimation error as the $p$-norm in the feature space:
\begin{equation}
    \error{\shortpred(v)}{\shortemb(v)}=\|\shortpred(v)-\shortemb(v) \|_p.
    \label{eq:estimation_error}
\end{equation}
For a set of nodes $V$, the mean estimation error will be denoted by $E(V)=(1/|V|)\sum\nolimits_{v\in V} \error{\shortpred(v)}{\shortemb(v)}$.

%%%%%%%%%%%%%%%%%%%%%%%%%%
\subsection{Node coherence}
\label{subsec:node_coherence}

Node coherence is a key concept in this article, and seeks to capture the degree of similarity of a set of nodes in the feature space.
Given $V\subset\Unodes$ for which features are known or estimated, we denote by $\incoherence(V)$ the incoherence of $V$.
Several incoherence metrics are possible, of which we propose to use the standard deviation of the distance to the centroid of $V$ in the feature space:
\begin{equation}
        \incoherence(V) = \sqrt{\frac{1}{|V|}\sum\limits_{v\in V}\| \shortemb(v)-c_V\|^2_p },
\label{eq:set_incoherence}
\end{equation}
\noindent{}where $c_V = (1/|V|)\sum\nolimits_{v\in V} \shortemb(v)$ is the centroid of $V$.

We prefer the definition of Equation \eqref{eq:set_incoherence} over other intuitive common metrics such as entropy \cite{onesto2019relating}: a set $V$ can have minimal entropy and high values of deviation $I(V)$ if the underlying probability distribution of $V$ is multi-modal in the feature space.

%%%%%%%%%%%%%%%%%%%%%%%%%%%%%%%%%%%%%%%%%
\subsection{$\varepsilon$-coherent neighborhoods of a set of nodes}

Whenever we have a set of nodes $V\subset\Unodes$, we can consider the up- or downstream neighborhoods of nodes that are also coherent: i.e., nodes that are linked to a set, and whose incoherence with respect to this set is below a given threshold $\varepsilon$. 
We formalize this notion, for both directions, as the \textit{$\varepsilon$-coherent upstream} and \textit{downstream neighborhood} of set $V$:

\begin{equation*}
n^\varepsilon_\dir(V) = \left\{ u\in n_\dir(V) : \incoherence(n_{\overline{\dir}}(u)\cap V)\leq \varepsilon \right\}
\end{equation*}

\noindent{}for $\dir\in\{\text{up},\text{down}\}$, and $\overline{\dir}$ the direction opposite to $\dir$.

\begin{figure}[h!]
    \centering
    \includegraphics[width=0.85\columnwidth]{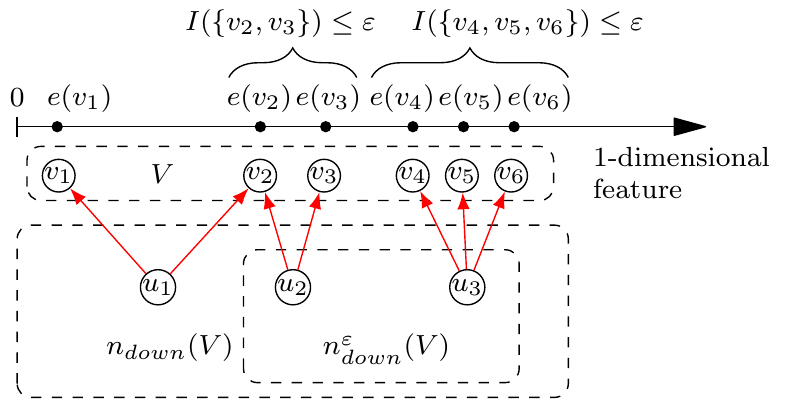}
    \caption{Illustration of the construction of a $\varepsilon$-coherent downstream neighborhood $n^\varepsilon_{\text{down}}(V)$ of a set $V$ embedded in a 1-dimensional feature space.}
    \label{fig:coherent_neighborhood}
\end{figure}

%%%%%%%%%%%%%%%%%%%%%%%%%%%%%%%%%%%%%%%%%%%%%%%%%%%%%%%%%%%%%%%%%%%%%%%%%%%%
\section{Propagation Methods for Latent Features in Social Networks}
\label{sec:value_propagation_methods}

We consider two different approaches for propagating ideological features in a larger set of the network: Method A) directed sequences of $\varepsilon$-coherent neighborhoods, and Method B) sequences of projections using $\varepsilon$-coherent neighborhoods.
Both families of methods are based on the concept of coherent neighborhoods, but motivated by distinct intuitions.
Method A assumes that the features attached to nodes are ``diffusing'' along the social network formed by follower/followee relationships. 
This method is rooted in the larger homophily hypothesis ``that birds of a feather flock together'' \cite{mcpherson2001birds}, meaning that two connected users are likely to demonstrate homophilic behavior, sharing a similar ideological position. 
Method B has different premises. 
Now, if unknown, ideological features of a node are estimated using those of other nodes that occupy a similar position in the network. 
This method is based on the structural equivalence hypothesis, which posits that two nodes sharing the same neighborhood are similar \cite{sailer1978structural}.

The approach of this second family of methods is, for example, at the core collaborative filtering approaches in algorithmic recommendation.
Here we will follow the same principle, estimating the ideological features of a user on Twitter, as the aggregation of features of users following, or being followed by the same users.

%%%%%%%%%%%%%%%%%%%%%%%%%%%%%%%%%%%%%%%
\subsection{Directed sequences of $\varepsilon$-coherent neighborhoods}
\label{subsec:definition_method_1}

This method generates two sequences of sets of nodes: one of coherent nodes $\{V_i\}_{i\geq 0}$, for which we estimate the ideology, and one of incoherent nodes $\{\overline{V}_i\}_{i\geq 0}$, that we avoid using in ideology estimation.

Let us consider a seed set $V_0\subset\Unodes$ for which features $\shortemb(v)$ for $v\in V_0$ are known, 
and a direction $\dir\in\{\text{up},\text{down}\}$. Starting at $V_0$, a directed sequence of $\varepsilon$-coherent sets of nodes $V_0,V_1,V_2,\ldots$ is computed as $V_{i+1} = V_i \cup \Delta V_{i}$ for $i=0,1,2,\ldots$, with

\begin{equation}
  \Delta V_{i} = \left\{v\in n^\varepsilon_{\dir} \left( V_{i} \right) : v\notin \left(V_i \cup \overline{V}_i \right) \right\}, 
 \label{eq:directed_coherent_sequence}
\end{equation}

\noindent{}where $\overline{V}_{i+1} = \overline{V}_i \cup \Delta \overline{V}_{i}$ for $i=0,1,2,\ldots$, with $\overline{V}_0=\varnothing$ and

\begin{equation}
\Delta \overline{V}_i = \left\{v\in n_{\dir}\left( V_{i} \right) : v\notin \left( \overline{V}_i \cup \Delta V_i \right) \right\}.
\label{eq:directed_incoherent_sequence}
\end{equation}

By definition, $\{V_i\}_{i\geq0}$ is an incremental ($V_{i}\subset V_{i+1}$) by disjoint additions ($V_i\cap\Delta V_i=\varnothing$) sequence of $\varepsilon$-coherent neighbors. 
Sequence $\{\overline{V}_i\}_{i\geq0}$ is also increasing ($\overline{V}_{i}\subset \overline{V}_{i+1}$) by disjoint additions ($\overline{V}_i\cap\Delta \overline{V}_i=\varnothing$), but of incoherent neighbors.
Nodes deemed incoherent are stored to avoid coherence collisions: an incoherent node cannot become coherent by virtue of new nodes who had been attributed features by propagation. 
Whenever new $\varepsilon$-coherent neighbors $\Delta V_i$ are discovered at the $i$-th step, the features of their nodes are estimated as 
\begin{equation}
    \shortpred(v) =\frac{1}{\left|n_{\overline{\dir}}(v)\cap V_i\right|} \sum\limits_{u\in n_{\overline{\dir}}(v)\cap V_i} \shortpred(u), \;\text{ for }v\in\Delta V_i,
\end{equation}

\noindent{}setting $\shortpred(v)=\shortemb(v)$ for $v\in V_0$.

\begin{figure}[h!]
    \centering
    \includegraphics[width=0.75\columnwidth]{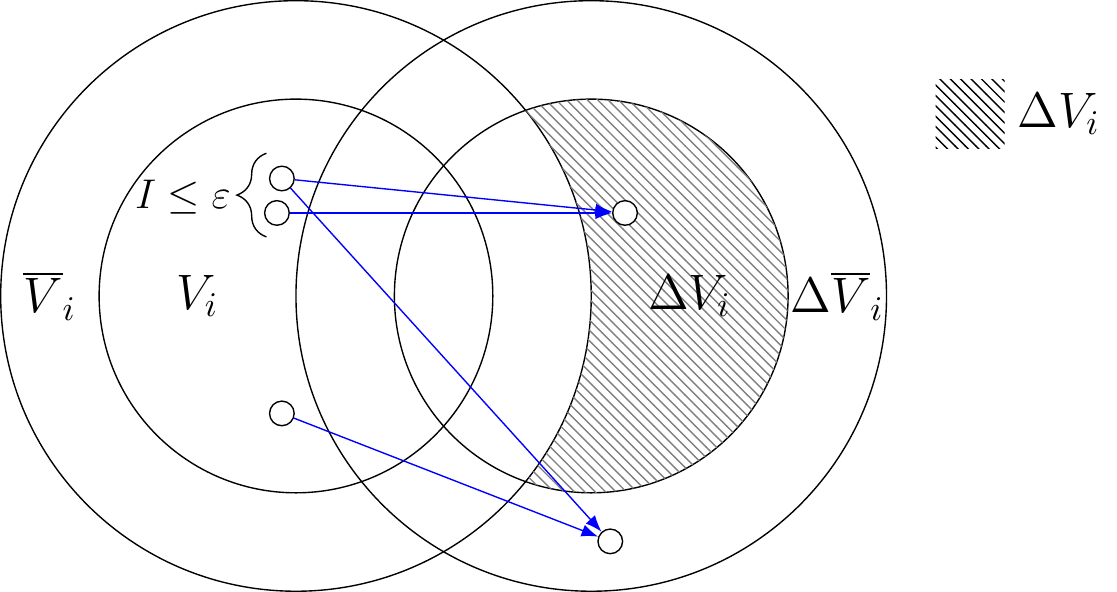}
    \caption{Schematic representation of the computation of $\Delta V_i$ and $\Delta \overline{V}_i$ on the $i$-th step of an upstream directed sequence of $\varepsilon$-coherent neighborhoods using \textbf{Method A}. In the upstream variant, users from $V_i$ follow users from $\Delta V_i$.}
    \label{fig:diagram_methodA}
\end{figure}

%%%%%%%%%%%%%%%%%%%%%%%%%%%%%%%%%%%%%%%
\subsection{Sequences of projections using $\varepsilon$-coherent neighborhoods}
\label{subsec:definition_method_2}

Inspired by the way in which features are computed in multi-dimensional scaling for users depending who they follow,  we propose a second family of methods, Method B, for feature propagation using $\varepsilon$-coherent neighborhoods. 
Most ideology scaling methods exploit the collective structural relations between nodes that are important for the latent features of a network (typically politicians) and their followers \cite{barbera2015birds,barbera2015tweeting}.
While intuitive, the first family of methods proposed in Section~\ref{subsec:definition_method_1} overlooks this aspect.
This can have disadvantages: for example, some nodes can attract followers without themselves following many nodes coherently. This could arguably be the case of Twitter accounts of media outlets, institutions, or public figures.
Accordingly, we propose a second family of methods centered around the notion of structural similarity: nodes that are similar because they follow (upstream) or are followed (downstream) by the same users, hereafter called \textit{pivots}.
As with the previous method, Method B generates two sequences of sets of nodes: one of coherent nodes $\{V_i\}_{i\geq 0}$, for which we estimate the ideology, and one of incoherent nodes $\{\overline{V}_i\}_{i\geq 0}$ that we now avoid using as pivots.

Let us consider a seed set $V_0\subset\Unodes$ for which features are known and a direction $\dir\in\{\text{up},\text{down}\}$.
Starting at $V_0$, a sequence of projected $\varepsilon$-coherent sets of nodes $V_0,V_1,V_2,\ldots$ is computed as $V_{i+1}=V_i\cup\Delta V_{i}$ for $i=0,1,2,\ldots$, for which we consider the set $\pivot^\varepsilon_i$ of $\varepsilon$-coherent pivots:
\begin{equation}
\pivot^\varepsilon_i = n^\varepsilon_\dir (V_i)\backslash{}\overline{V}_i,
\end{equation}

\noindent{}where $\overline{V}_{i+1}=\overline{V}_i \cup \Delta\overline{V}_i$ for $i=0,1,2,\ldots$, with $\overline{V}_0=\varnothing$, and

\begin{equation}
    \Delta \overline{V}_i = \left\{v\in n_{\dir}(V_i): v\notin\left( n^\varepsilon_{\dir}(V_i)\cup \overline{V}_i \right) \right\}.
\end{equation}

As with the previous family of Method A, the sets $\overline{V}_i$ store the nodes deemed incoherent and that cannot be used, but now  as pivots.
The sets $\pivot^\varepsilon_i$ are used at each iteration to compute additions $\Delta V_i$ according to coherent structural  similarity:

\begin{equation}
    \Delta V_i = \left\{v\in n^\varepsilon_{\overline{\dir}}\left( \pivot^\varepsilon_i \right): v \notin V_i \right\}.
\end{equation}

Again by definition, sequences $\{V_i\}_{i\geq0}$ and $\{\overline{V}_i\}_{i\geq0}$ are incremental by disjoint additions.
In contrast with the first family of Method A from Section~\ref{subsec:definition_method_1}, now it is the coherence of the pivot nodes in sets $\pivot^\varepsilon_i$ that is assured.

Whenever new neighbors $\Delta V_i$ are discovered at the $i$-th iteration, their features are estimated as 
\begin{equation}
    \shortpred(v) =\frac{1}{\left|\coneighbors_i(v,\pivot^\varepsilon_i,V_i)\right|} \sum\limits_{u\in \coneighbors_i(v,\pivot^\varepsilon_i,V_i)} \shortpred(u), \;\text{ for }v\in\Delta V_i,
\end{equation}
\noindent{}where $\coneighbors_i(v,\pivot^\varepsilon_i,V_i)$ is the set of co-neighbors of $v$ in $V_i$ through pivot  $\pivot^\varepsilon_i$:
\begin{equation}
    \coneighbors_i(v,\pivot^\varepsilon_i,V_i)= V_i \cap n_{\overline{\dir}}\left(  n_{\dir}(v) \cap \pivot^\varepsilon_i   \right) .
    \label{eq:coneighbors}
\end{equation}

\begin{figure}[h!]
    \centering
    \includegraphics[width=0.7\columnwidth]{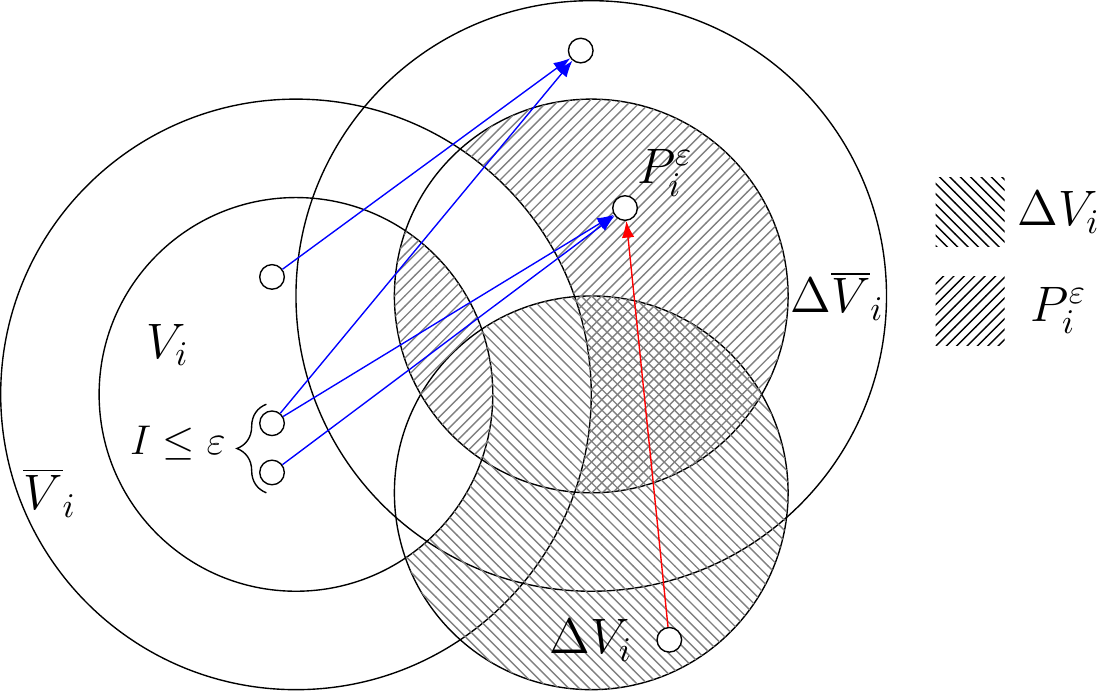}
    \caption{Schematic representation of the computation of $\Delta V_i$ and $\Delta \overline{V}_i$ using pivot $\pivot^\varepsilon_i$ on the $i$-th step of an upstream (co-followers-based) sequence of projected $\varepsilon$-coherent neighborhoods using \textbf{Method B}. In the upstream variant, users from $V_i$ and $\Delta V_i$ co-follow users from pivot set $\pivot^\varepsilon_i$.}
    \label{fig:diagram_methodB}
\end{figure}

%%%%%%%%%%%%%%%%%%%%%%%%%%%%%%%%%%%%%%%%%%%%%%%%%%%%%%%%%%%%%%%%%%%%%%%%%%%%
\section{French Political Twitter Dataset}
\label{sec:illustrative_case_data}

In order to analyze the feature propagation methods described in the previous section on real data, here we present a pertinent part of the Twitter network\footnote{In conformity with the General Data Protection Regulation 2016/679, the project called \textit{Ideology scaling Twitter France}, whose data we exploit, has been declared the 19 Mars 2020 at the registry of data processing at the \textit{Fondation Nationale de Sciences Politiques} (Sciences Po), and respects Twitter's data use policies.} and an ideological scaling producing an interpretable ideological feature space.

%%%%%%%%%%%%%%%%%%%%%%%%%%%%%%%%%%%%%%%%%%%
\subsection{French MPs and their followers}

Our data collection starts with the set $\MPset$ of the $\setPsize$ (out of 925) French MPs present on Twitter\footnote{The list of French MPs present on social networks is provided by the French chambers of parliament at \url{http://www2.assemblee-nationale.fr/deputes/liste/reseaux-sociaux}, for deputies, and \url{http://www.senat.fr/espace_presse/actualites/201402/les_senateurs_sur_twitter.html} for senators.}, belonging to 10 main different parties or groups/alliances of parties.
We then proceeded to collect all the followers of the accounts of MPs in $\MPset$. 
This collection was conducted on May 2019, and resulted in the constitution of the set $\MPfol=n_\text{down}(\MPset)$ of followers of $\MPset$, which amounts to $|\MPfol|=\setFsize$ unique Twitter accounts.

%%%%%%%%%%%%%%%%%%%%%%%%%%%%%%%%%%%%%%%%%%%
\subsection{An ideological inference procedure via multidimensional scaling}

We follow the methodology described by Barbera et al. \cite{barbera2015tweeting} for inferring ideological features through multidimensional scaling of the sub-graph of the MPs and their followers.
After removing from this sub-graph the followers that follow less than 3 MPs, and then removing users that had a repeated set of followed MPs (to assure full rank of the adjacency matrix), we obtained a set $\MPfilfol$ of $\setFtsize$ accounts.
We represent this sub-graph as a $\{0,1\}^{|\MPfilfol|\times|\MPset|}$ adjacency matrix , where a relation of following is encoded with the value 1, and its absence with the value 0.
Next, we produce a reduced-dimensionality representation of these $\setFtsize$ observations using a Correspondence Analysis (CA) \cite{greenacre2017correspondence}.
The first 2 principal components of the reduced-dimensionality space, PC1 and PC2, explain 0,97\%{} and 1,07\%{} of the inertia.
This might seem as little explanation of the variability in the observed ways in which users from $\MPfilfol$ follow MPs.
However, the experience in several countries has revealed that the first components offer useful interpretations in the Twitter MPs' network.
Examples include the UK, Spain, Italy, and Netherlands \cite{barbera2015birds}).
Similar results have also been found in France \cite{briatte2015recovering,cointet2020color}.
In accordance with analyses made for the cases of other countries, those of France suggest feature interpretation where the first component is identified with left-right ideology.

\subsection{Interpreting the ideological feature space }

\begin{figure}
\centering
\includegraphics[width=\columnwidth]{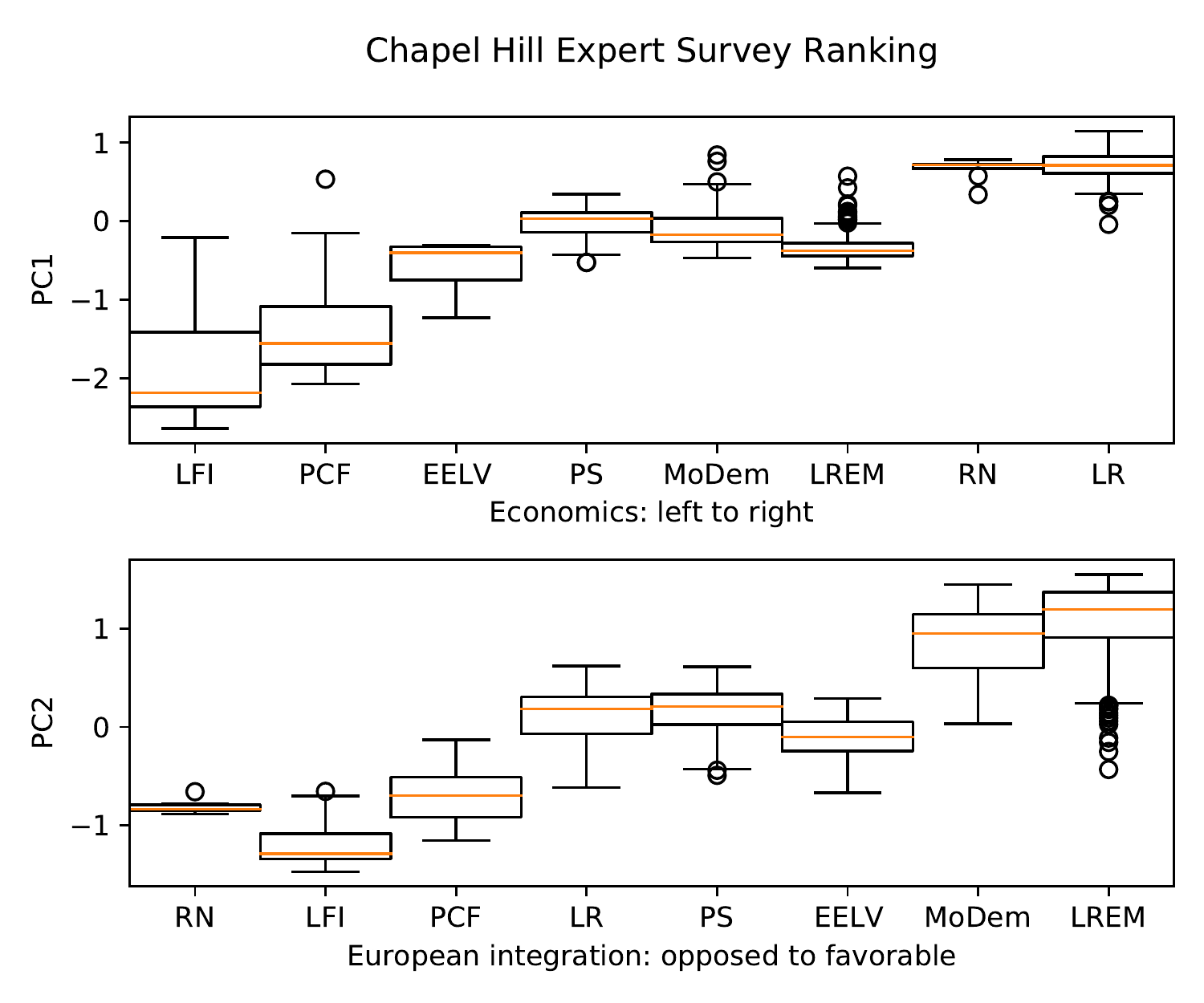}
\caption{Distribution of the embedded ideological features PC1 and PC2 for the French MPs on Twitter per party. Parties are ordered according to the two criteria using the 2019 Chapel Hill Expert Survey data: left-right economics for feature PC1 (top), and attitudes toward European integration for feature PC2 (bottom).}
\label{fig:boxplot_embedded_values_french_MPs}
\end{figure}

To provide and interpretation of PC1 and PC2 for our sets $\MPset$ and $\MPfilfol$, we consider exogenous variables for the parties of the MPs.
We use the 2019 Chapel Hill Expert Survey (CHES) data  \cite{chesdata2019}: an estimation on ideology and policy position of European parties, made by 421 political scientists specializing in political parties.
Out of the 10 political parties identified for accounts in $\MPset$, 8 are also present in the CHES data.
We computed the correlations between the positions of MPs on PC1 and PC2 of the feature space with all the 51 criteria were CHES data produces estimation for their political parties.
This allowed to identify the two most relevant criteria related to our axes.
Fig.~\ref{fig:boxplot_embedded_values_french_MPs} shows the ordering of these eight parties according to the two different criteria: 1) parties' economic views, from left to right, and 2) parties' attitudes towards European integration, from opposed to favorable. For each party in Fig.~\ref{fig:boxplot_embedded_values_french_MPs}, we provide the embedded features of PC1 per MP for criterion 1 (left-right ideology), and the embedded features of PC2 per MP for criterion 2 (attitudes towards European integration).
While it is not the objective of this article, 
Figure~\ref{fig:boxplot_embedded_values_french_MPs} illustrates the potential applicability of the multidimensional scaling procedure.

\begin{figure}
\centering
\includegraphics[width=\columnwidth]{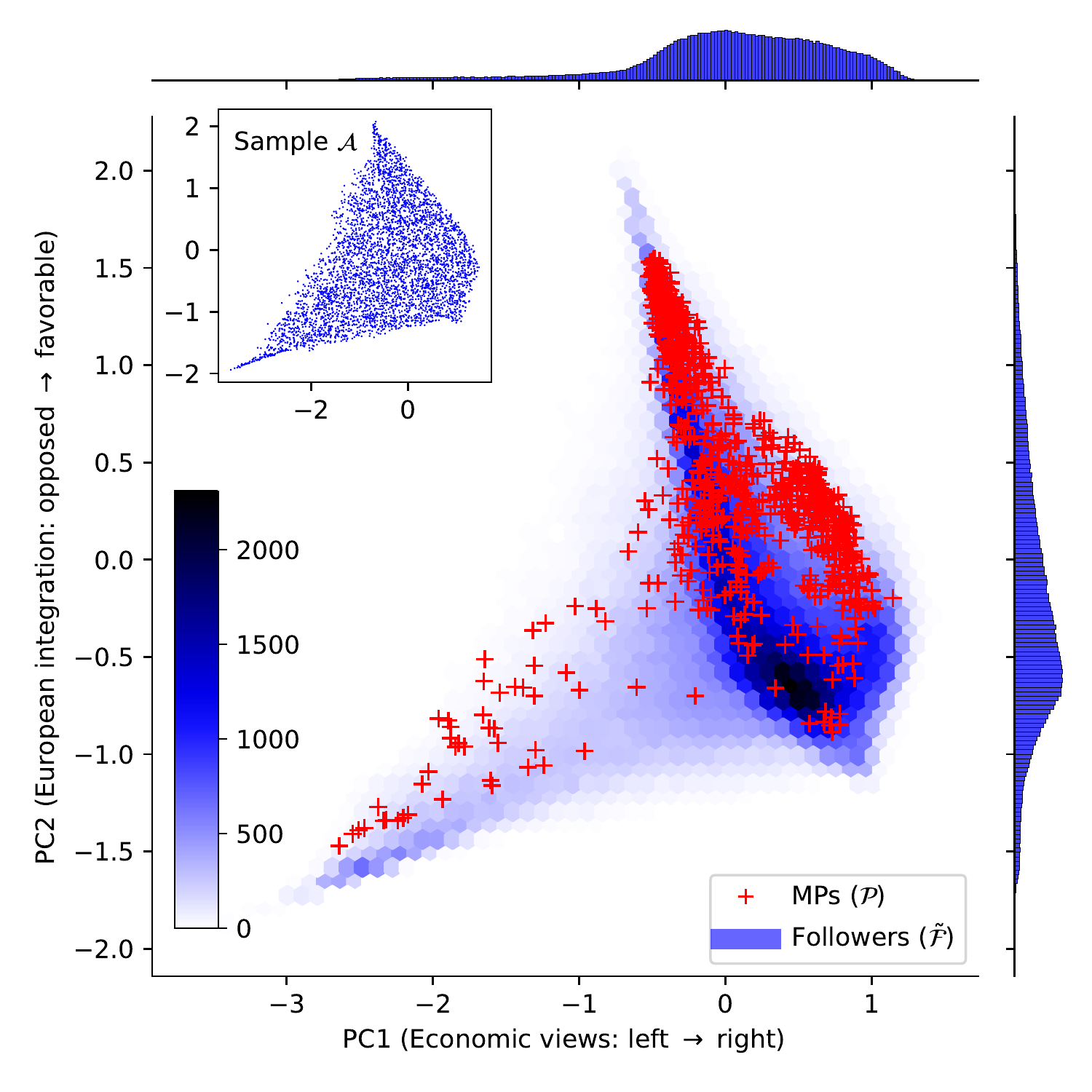}
\caption{Position of $\MPset{}$ French MPs in the first two embedded dimensions, PC1 (left to right ideology) and PC2 (pro- and against European integration), histogram of the positions of their followers $\MPfilfol{}$, and positions of sampled accounts in set $\Aset$.}
\label{fig:2d_histogram_embedded_seed_bipartite}
\end{figure}

Under this interpretation, the French Twitter network of MPs and their followers (the sets $\MPset$ and $\MPfilfol$) may be represented in a 2-dimensional latent ideological feature space.
In this feature space, the first dimension (PC1) provides an index for the concept of \textit{left} and \textit{right} positions, and the second dimension (PC2) provides an index for attitudes towards European integration, which is in line with observations that have been made in the case of other countries \footnote{Research results in several countries have also identified this second axis as related with issues dividing public along pro- and anti-establishment positions \cite{schmidt2020measuring,chiche2000espace}.}.
Fig.~\ref{fig:2d_histogram_embedded_seed_bipartite} illustrates the positions of the sets $\MPset$ of parliamentarians and $\MPfilfol$ of their followers in
this bidimensional ideological feature space. 

%%%%%%%%%%%%%%%%%%%%%%%%%%%%%%%%%%%%%%%%%%%
\subsection{Datasets for experimentation}

We are now concerned with the problem of using the set $\MPfilfol$ with $\setFtsize$ accounts with known features to establish datasets for the testing and evaluation.
We cannot set $V_0=\MPfilfol$ for testing and evaluation.
This is because when producing the next set $V_1$ in the sequence (with either method) the new added nodes would not have known \textit{true} features $\shortemb(v)$ (as computed with ideological scaling) to test the estimated ones $\shortpred(v)$, for $v\in \Delta V_0$.
To circumvent this difficulty, we take a subset $\Aset\subset\MPfilfol$, and then collect its upstream and downstream neighborhoods to use in the described methods, allowing for some elements of these new neighborhoods to be also in $\MPfilfol$.
To account for the possible specificities of the different regions of the ideological feature, we sample $\setAsize$\footnote{Originally, we sampled 5.000 nodes, but the later collection of followers and followees for 517 of these nodes was not possible because these accounts were either closed between collections, or private and information could not be collected.} nodes uniformly in space from $\MPfilfol$ to produce set $\Aset$ (see Fig.~\ref{fig:2d_histogram_embedded_seed_bipartite}).
Had we not sampled with spatial uniformity, the sampling would have produced a majority of nodes near the origin of the ideological feature space (see the concentration of nodes near the origin in  Fig.~\ref{fig:2d_histogram_embedded_seed_bipartite}).
Next, we collect the followees/friends of $\Aset$ as $\Bset = n_{\text{up}}(\Aset)$ (obtaining $|\Bset|=\setBsize$), and the followers of set $\Aset$ as $\Cset = n_\text{down}(\Aset)$ (obtaining $|\Cset|=\setCsize$).
This sub-sampling and collection operations achieve sets for our test evaluations that are such that $|\Bset \cap \MPfilfol| = \setFinterB$, and $|\Cset \cap \MPfilfol| = \setFinterC$.

%%%%%%%%%%%%%%%%%%%%%%%%%%%%%%%%%%%%%%%%%%%%%%%%%%%%%%%%%%%%%%%%%%%%%%%%%%%%
\section{Numerical Experiments}
\label{sec:numerical_results}

In this section we analyze some results of the methods proposed in Section~\ref{sec:value_propagation_methods} for estimating ideological features on Twitter data. 
For both methods, we center the analysis around the set $\Aset$ built for this purpose, and described in the previous section.
We will be interested in analyzing, for different values of coherence $\varepsilon$, 1) the accuracy (the degree of error in estimating ideological features), and 2) the coverage (the size of the disjoint additions with estimated features) of the proposed methods.
The main aspect we seek to investigate in this section is the relation between the coherence threshold parameter $\varepsilon$ and the trade-off between estimation error and size of a disjoint addition $\Delta V_i$ at a step $i$. 
Throughout this section, we compute error $E$ (cf. Equation~\eqref{eq:estimation_error}) and coherence $\incoherence$ (cf. Equation~\eqref{eq:set_incoherence}) using $p=2$.

%%%%%%%%%%%%%%%%%%%%%%%%%%%%%%%%%%%%%%%%%%%%%
\subsection{Directed sequences of $\varepsilon$-coherent neighborhoods}

\begin{figure}[h!]
     \centering
     \includegraphics[width=\columnwidth]{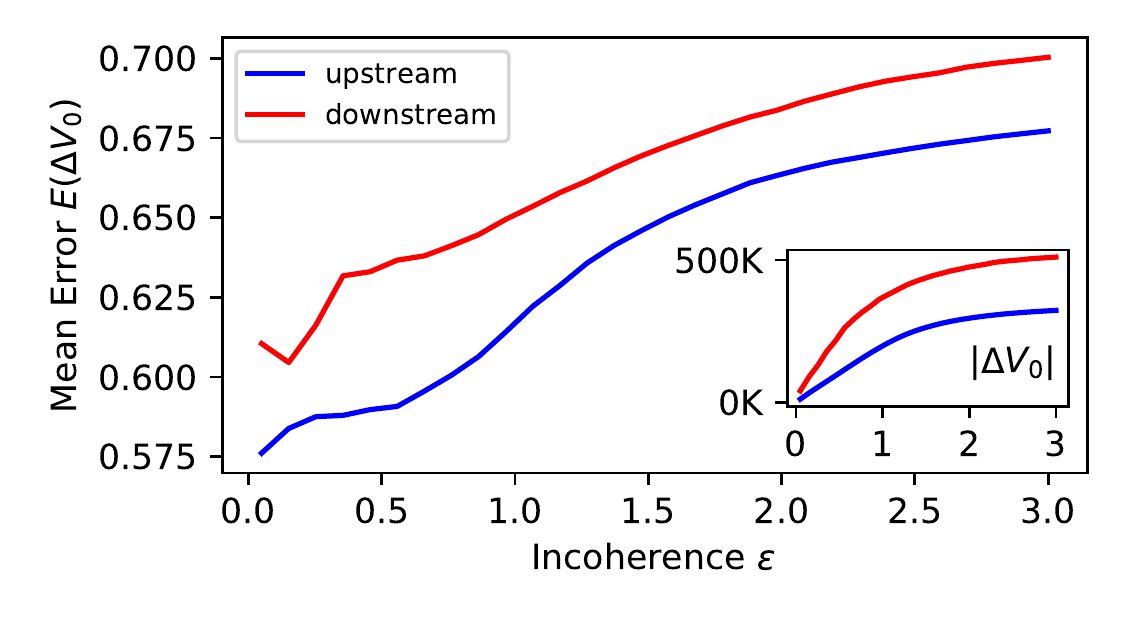}
     \caption{Mean estimation error $E(\Delta V_0)$ and size of $\Delta V_0$ for Method A in the upstream and downstream directions, using $V_0=\Aset$.}
     \label{fig:1st_iterations_directed_sequence}
 \end{figure}

Because we have collected the upstream and downstream neighborhoods of set $\Aset$ (sets $\Bset$ and $\Cset$), we can readily compute the first step for Method A in both directions.
Figure~\ref{fig:1st_iterations_directed_sequence} reports the error $E(\Delta V_0)$ and the size of $\Delta V_0$ for the first step ($i=0$).

While intuitive, the estimation of ideological features through directed propagation of Method A in either direction performs poorly. 
Imposing higher levels of coherence (lower thresholds $\varepsilon$ for incoherence) does improve the mean estimation error of nodes in 
$\Delta V_0$.
However, this improvement is marginal and, most importantly, limited.
Upstream neighbors (friends/followees) of users in $\Aset$ that are followed by highly coherent users, have their ideology estimated with a mean error no lower than $E(\Delta V_0)=$0.6, which amounts to 12.5\%{} of the diameter of $\MPfilfol$ in the feature space.
Similarly, downstream neighbors (followers) of users in $\Aset$ that follow highly coherent users, have their ideology estimated with a mean error no lower than $E(\Delta V_0)=$0.575 (12.0\%{} of the diameter of $\MPfilfol$).
Had set $\Delta V_1$ be computed (upstream or downstream), its ideological features would be estimated only from those already --poorly-- estimated for set $\Delta V_0$ (disjoint additions, assured by Equation~\eqref{eq:directed_coherent_sequence}, imply that $n_{\dir}(V_0) \cap \Delta V_1 = \varnothing$).

%%%%%%%%%%%%%%%%%%%%%%%%%%%%%%%%%%%%%%%%%%%%%
\subsection{Sequences of projections using $\varepsilon$-coherent neighborhoods}

\begin{figure}[h!]
     \centering
     \includegraphics[width=\columnwidth]{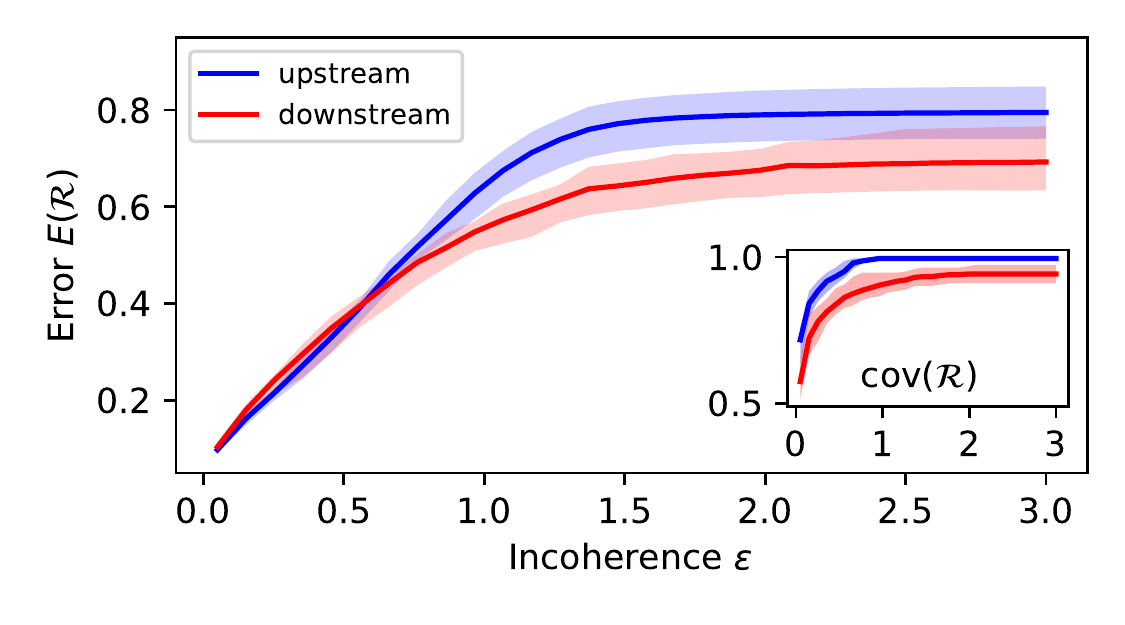}
     \caption{Estimation error and coverage of set $\result=\Aset^K_\text{test} \cap \Delta V_0$ and $P_0$ for the Method B in a $K$-fold cross-validation retrieval of ideologies of $\Aset$ ($K=$20). The distribution of errors and coverage over the folds is reported with the median, the minimum, and the maximum  values .}
     \label{fig:1st_iteration_projected_sequence}
 \end{figure}

The setting of Method B is similar to that of collaborative filtering (CF) \cite{schafer2007collaborative} in Recommender Systems.
In user-based CF, the similarity of users is computed according to structural similarity on a bipartite graph: users are similar if they have chosen similar items.
Items can then be proposed to a user among those already chosen by similar users (structural similarity is related to preference).
In the setting of our Method B, users are deemed similar if they follow, or are followed by (depending if the selected direction is upstream or downstream) similar users of a so-called pivot set for a predetermined coherence $\varepsilon$.
Taking on known evaluation protocols for CF \cite{bobadilla2012collaborative}, we propose a method for assessing the accuracy and coverage of Method B.
Given a positive integer $K$, we perform a $K$-fold bipartite cross-validation. We divide $\Aset$ in $K$ parts, taking one as $\Aset^K_\text{test}$ and the rest as $\Aset^K_\text{train}$.
We set $V_0 = \Aset^K_\text{train}$, we compute $P^\varepsilon_0$ using $\Bset$ and $\Cset$ for upstream and downstream directions, and then set $\result=\Aset^K_\text{test} \cap \Delta V_0$.
To analyze the coverage, we examine the quantity $\text{cov}(\result) = |\result|/|\Aset^K_\text{test}|$, ranging from 0 (no coverage) to 1 (total coverage).
We choose $K=20$ and report the results for varying values of incoherence $\varepsilon$, providing, for the ensemble of the 20 folds, the median, the maximum, and the minimum value in Fig.~\ref{fig:1st_iteration_projected_sequence}.

Results obtained with Method B, reported in Fig.~\ref{fig:1st_iteration_projected_sequence}, are more satisfactory than those obtained with Method A. 
Thresholds for incoherence below  $\varepsilon=1$ (approximately) already allow for less error in estimation in comparison to Method A.
In fact, under our setting, estimation errors for Method B can be made as low as (approximately) 0.1 in the feature space by sacrificing coverage of new nodes added in the following step of the sequence.
With a small value of threshold for $\epsilon$ (lower than 0.1) the estimation error is around 0.1 in distance in the feature space (2.1\%{} of the diameter of $\MPfilfol$).
This trade-off in coverage does not come at great expense, as suggested by our proposed metric $\text{cov}(\result)$: at least 50\%{} of nodes left in the test set $\mathcal{A}^K_{\text{test}}$ can be recovered, independently of the threshold for incoherence.
The amount of nodes in the set of pivots $\pivot^\varepsilon_0$ is that of $\Delta V_0$ for Method A, and follows intuition in that, when forcing less incoherence, less nodes are available for the search of co-neighbors (cf. Equation~\eqref{eq:coneighbors}).

While it was possible to evaluate the quality of Method A precisely, the same evaluation would have required the collection of the downstream neighbors of $\Bset$ (i.e. $n_{\text{down}}\left( n_{\text{up}}(\Aset)\right)$) and the upstream neighbors of $\Cset$ (i.e. $n_{\text{up}}\left( n_{\text{down}}(\Aset)\right)$), which would amount to a mixed-direction snowball-sampling of the Twitter network from $\Aset$.
Thus the need for a different evaluation protocol for both methods.
While there is a slight risk of over-estimating the coverage capacity of Method B using our testing protocol (due to the fact that $\result$ is restricted to $\Aset$) it has the great advantage of allowing for experimentation with large sets: the (up- and downstream) neighbors of $\Aset$, i.e., the total population for feature propagation, account for nearly 6.5M users.

%%%%%%%%%%%%%%%%%%%%%%%%%%%%%%%%%%%%%%%%%%%%%%%%%%%%%%%%%%%%%%%%%%%%%%%%%%%%
\section{Conclusions}
\label{sec:conclusions}

We proposed two methods (A and B) for the propagation of latent ideological features on Twitter.
These methods use propagation for the estimation of ideology of users, represented in a continuous n-dimensional features space.
Method A is based on the homophily hypothesis: if users are connected (one follows the other) their ideology is similar.
Method B is based on the structural similarity hypothesis: if users are connected to the same neighborhoods their ideology is similar.
In these methods, we proposed the consideration and formalization of these notions when depending on ideological coherence of neighbors, which we model with a single parameter.

To analyze our methods, we collected Twitter data using the accounts of French MPs.
We were able to produce a 2-dimensional ideological embedding for a sub-graph of nearly 370K users. With the help of the Chapel Hill Expert Survey, we validate the interpretation of the two emerging dimensions as related to left-right ideology, and attitudes toward European integration.
Collecting again neighborhoods of some of these users we used our methods to propagate these ideological features with predefined degrees of accuracy within a potential population of nearly 6.5M users.
To total number of users for which ideology is propagated, depends on the error to be accepted, which is determined by a coherence threshold $\varepsilon$.
The protocols established for evaluation allow us to analyze the relation between estimation accuracy for ideology in propagation, and the coverage of the method (i.e., the number of users to which we can propagate).
The trade-off between these two competing and desirable properties was modulated by our coherence parameter.
Analyzing Methods A and B, we find that, when coherence is considered, the ideology of a user on Twitter is better estimated using other users that are structurally similar (Method B), than using other users that might be directly connected (Method A).
These results lend support to the structural similarity hypothesis over the homophily hypothesis.
Even more: while the difference is small, the ideology of a user is better estimated using users that follow the same accounts, than using users that are followed the same Twitter accounts.

The formalism used to treat coherence in social networks, and the methods proposed for propagation, can be extended to any other network were directed edges model the fact a user can receive information from another.
On Facebook, for example, the fact that user $a$ and user $b$ are \textit{friends}, could be represented by two directed edges between both users, and in different directions.
In other networks, such as Instagram or YouTube (using the \textit{subscription} relation) the formalism and methods can be directly applied.

%%%%%%%%%%%%%%%%%%%%%%%%%%%%%%%%%%%%%%%%%%%%%%%%%%%%%%%%%%%%%%%%%%%%%%%%%%%%

\section*{Acknowledgments}

We thank Justin Clarke and Benjamin Ooghe-Tabanou for their help in the collection of Twitter data.

%%%%%%%%%%%%%%%%%%%%%%%%%%%%%%%%%%%%%%%%%%%%%%%%%%%%%%%%%%%%%%%%%%%%%%%%%%%%
\bibliographystyle{IEEEtran}
\bibliography{references}

\end{document}